\documentclass[twocolumn,astrosymb]{aastex631}
\usepackage{amsmath,amssymb}
\usepackage{subfigure}
\usepackage{xcolor}
\usepackage[T1]{fontenc}

\usepackage[english]{babel}
\usepackage{graphicx}
\usepackage{wrapfig}

\graphicspath{{./}{figures/}}

\begin{document}

\title{The Not-So Dramatic Effect of Advective Flows on Gas Accretion}

\correspondingauthor{Vincent Savignac}
\email{vincent.savignac@mail.mcgill.ca}

\author[0000-0002-2870-6940]{Vincent Savignac}
\affiliation{Department of Physics and Trottier Space Institute, McGill University, 3600 rue University, H3A 2T8 Montreal QC, Canada}
\affiliation{Trottier Institute for Research on Exoplanets (iREx), Universit\'e de Montr\'eal, Canada}

\author[0000-0002-1228-9820]{Eve J. Lee}
\affiliation{Department of Physics and Trottier Space Institute, McGill University, 3600 rue University, H3A 2T8 Montreal QC, Canada}
\affiliation{Trottier Institute for Research on Exoplanets (iREx), Universit\'e de Montr\'eal, Canada}

\begin{abstract}

Super-Earths and mini-Neptunes are the most common types of exoplanets discovered, yet the physics of their formation are still debated. Standard core accretion models in gas-rich environment find that typical mini-Neptune mass planets would blow up into Jupiters before the underlying disk gas dissipates away. The injection of entropy from the protoplanetary disk into forming gaseous envelopes has recently been put forward as a mechanism to delay this runaway accretion, specifically at short orbital distances. Here, we reevaluate this line of reasoning by incorporating recycling flows of gas into a numerical one-dimensional thermodynamic model with more realistic equation of state and opacities and the thermal state of the advective flow. At 0.1 AU, we find that advective flows are only able to produce mini-Neptunes if they can penetrate below $\sim$0.25 of the planet's gravitational sphere of influence. Otherwise, the gas-to-core mass ratio (GCR) reaches above $\sim$10\% which is too large to explain the measured properties of mini-Neptunes, necessitating other gas-limiting processes such as late-time core assembly. The effect of entropy advection on gas accretion weakens even further beyond 0.1 AU. We present an updated scaling relation between GCR and the penetration depth of the advective flows which varies non-trivially with orbital distances, core masses and dusty vs.~dust-free opacity. We further demonstrate how measurements of planet mass distribution beyond $\sim$1 AU using future instruments such as the Nancy Grace Roman Space Telescope could be used to disambiguate between different formation conditions of gas-poor planets.

\end{abstract}


\section{Introduction} \label{Intro}
Super-Earths and mini-Neptunes dominate the discovered exoplanetary populations, with $\sim$30--50\% of all Sun-like stars harboring at least one of these planets within orbital periods of $\sim$300 days \citep[e.g.,][]{Howard,Batahla,Petigura_2013,Dong_2013,Fressin,Rowe,Burke_2015,Zhu_2018}. The mini-Neptunes in particular have measured masses of $\sim$2-20 $M_\oplus$ \citep{Weiss_Marcy,Wu_Lithwick} and radii of $\sim$1.8--4$R_\oplus$, which imply a thin H-He atmosphere atop a rocky core with the envelope mass fraction ranging from $\sim$1 \% to 10\% \citep{Rogers_2010, Lopez_2014,Wolfgang_2015}. Classical theories of core accretion expected such planets (especially the more massive ones $\gtrsim$10$M_\oplus$) to have undergone a runaway gas accretion and blow up into gas giants \cite[e.g.,][]{Mizuno,Stevenson,Pollack,Ikoma}, prompting a need for a revision to such a theory to explain the existence and prevalence of mini-Neptunes.

One solution is to consider the late-time assembly of planetary cores which delays the onset of gas accretion \citep{Lee_2014,Lee_2016}. By comparing the empirically determined orbit-crossing timescale from \cite{Zhou_2007} with the eccentricity damping timescale from gas dynamical friction \citep[e.g.,][]{Papaloizou_2000}, \citet{Lee_2016} argued that the nebular gas needs to be depleted by about four orders of magnitude with respect to a solar-composition minimum-mass nebula for the protocores to merge. By coupling a direct N-body simulation with gas accretion, \cite{Dawson} demonstrated that the core mergers can begin from the gas depletion factor of 1000, and \cite{Choksi} identified the peaks in the observed orbital period ratio of multi-planetary systems just wide of mean motion resonance can be explained by eccentricity damping and short-scale migration in heavily gas-depleted environment, by 3--5 orders of magnitude with respect to solar nebula, in agreement with previous literature. Such gas-depleted environment is consistent with the very late stage of disk evolution and so the limited time and the lower nebular density prevents the runaway accretion and ensures the accumulation of $\sim$1--10\% by mass envelope \citep{Lee_2016}.

An alternative mechanism to delay runaway is to consider hydrodynamic effects such as planetary rotation \citep{Zhong_2021} and the injection of entropy from the outer disk inside the envelope by ``atmospheric recycling'' (\citealt{Ormel_2015b}; see also \citealt{Fung_2015}). Three-dimensional (3-D) hydrodynamic models report the advective flows can penetrate deep into the envelope which can slow down the cooling of the planet as these flows are expected to bring in the (higher) disk entropy into the deep envelopes and therefore effect the formation of mini-Neptunes instead of Jupiters \citep{Lambrechts,Popovas_2018,Bethune_2019,Moldenhauer_2021,Bailey_2023}. Some of the the earliest studies \citep[e.g.,][]{Ormel_2015b} limit their considerations to isothermal gas flow which leads to no bound gas at all implying some underlying issue in the characterization of the flows. \cite{Ali_Dib_2020} accounted for this atmospheric recycling in one-dimensional (1-D) semi-analytical thermal calculations by modeling the upper advective region as an adiabat since they argue that the advection timescale is short compared to the cooling timescale of the envelope. They found that at short orbital distances, the effect is strong enough to significantly stall accretion which could explain why mini-Neptunes do not undergo runaway and remain small.

In their comparison between 1-D semi-analytical thermal calculations and 3-D global radiative simulations that include realistic opacities and equation of state, \cite{zhu_2021} found the upper advective region to be {\it not} an adiabat. They further found that the thermal state of the envelope should not be significantly altered by recycling at least at 5 AU which is where they focused their simulations \citep[see also][]{D'Angelo_2013,Ormel_2015b}. As high entropy flows from the protoplanetary disk try to penetrate a low entropy region of the envelope, buoyancy forces prevent the flows from penetrating deep in the atmosphere, as is commonly found in more realistic non-isothermal simulations \citep{Kurokawa}. \citet{zhu_2021} note however that it may still be the case that such advective flows could potentially inhibit runaway accretion at short orbital distances ($\sim$0.1 AU) where \cite{Moldenhauer_2021} argue that recycling flows can penetrate all the way to the bottom of the envelope. 

In this work, we revisit the role of recycling at short orbital distances by accounting for the outer hydrodynamic flow in 1-D thermal calculations. Although our approach is similar to the one proposed by \cite{Ali_Dib_2020}, the main difference is that we do not treat the recycling flows as an adiabatic process. Instead, we favor a self-consistent treatment of the advective flow accounting for its cooling while also using realistic equation of state and opacities. We ultimately seek to quantify the effect of entropy advection on its ability to slow down gas accretion and whether it can successfully produce mini-Neptunes at short orbital periods, even in gas-rich environments.

The paper is organized as follows. We outline the construction of envelope profiles and their time evolution in Section \ref{Methods}. Results are presented in Section \ref{results} along with an updated semi-analytic scaling relationship between the envelope mass fraction, time, and the depth of the advective flow. Section \ref{Discussion} answers whether or not entropy advection can explain mini-Neptunes at short orbital distance on its own, discusses the importance of using realistic opacity and equation of state in thermal evolution of planetary envelopes, summarizes how mini-Neptunes avoid runaway in the context of our results, and presents the critical core mass that separates gas-poor from gas-rich planets over a range of orbital distances and avenues for identifying likely formation conditions of gas-poor planets using future instruments such as the Nancy Grace Roman Space Telescope. Finally, we summarize and conclude in Section \ref{Conclusion}.

\section{Time-dependent model atmospheres} 
\label{Methods}

We follow the spherically symmetric model of \cite{Lee_2014} to simulate the formation of an atmosphere around a rocky core while also incorporating atmospheric recycling into the calculations. We first build a series of ``hydrostatic snapshots" of the envelope, each corresponding to different gas-to-core mass ratio GCR $\equiv M_{\rm gas}/M_{\rm core}$. These snapshots are threaded together by computing the time it takes for the envelope to cool from one snapshot to the next.

\subsection{Hydrostatic snapshots}
\label{equation_system}
For each value of GCR, we solve the standard stellar structure equations:

\begin{equation}
    \label{dM}
    \frac{dM(<r)}{dr}=4\pi r^2 \rho
\end{equation}

\begin{equation}
    \label{dP}
    \frac{dP}{dr}=-\frac{GM(<r)}{r^2}\rho -\frac{GM_{*}r}{a^3}\rho
\end{equation}

\begin{equation}
    \label{dT}
    \frac{dT}{dr}=\frac{T}{P} \frac{dP}{dr} \nabla
\end{equation}
for the density $\rho$, the pressure $P$, the temperature $T$, and the enclosed mass $M(<r) \equiv M$ as functions of the radius $r$. Here, $G$ is the gravitational constant, $M_\star$ is the mass of the central star which we fix to the solar mass, and $a$ is the orbital distance from the star. The term on the far right of Equation (\ref{dP}) is inserted to account for the central star's gravitational field as suggested by \cite{Zhu_2022}. The effect of this correction is small as we expect $M/r^3 \gg M_* / a^3$ within the atmosphere. The dimensionless temperature gradient $\nabla \equiv d$ ln $T/d$ ln $P$ depends on whether the energy transport is dominated by radiation or convection. Convection initially dominates throughout the envelope, but as the gas cools and more mass is accreted, radiation zones develop in the outer regions of the envelope. From radiative diffusion,

\begin{equation}
    \label{grad_rad}
    \nabla_{\rm rad} = \frac{3\kappa P}{64\pi GM\sigma T^4}L
\end{equation}
where $\kappa$ is the opacity, $L$ is the internal luminosity of the envelope, and $\sigma$ is the Stefan-Boltzmann constant. When the energy transport is dominated by convection,

\begin{equation}
    \nabla_{\rm ad} = - \frac{\partial \mathrm{log} S}{\partial \mathrm{log}P}
    \Big |_T \Big (\frac{\partial \mathrm{log} S}{\partial \mathrm{log}T}
    \Big |_P \Big ) ^{-1}
\end{equation}
where  $S$ is the specific entropy of the gas. 

We relate the state variables ($P$, $T$, $\rho$ and $S$) in the atmosphere with the equation of state (EOS) computed in \cite{Lee_2014} which includes H (molecular, atomic, and ionized), He, and metallic species in solar elemental abundances \citep{Grevesse_1993}. We fix mass fractions of X=0.7 for H, Y=0.28 for He and Z=0.02 for metals. As we will discuss more in Section \ref{Discussion}, adopting this realistic equation of state instead of a fixed adiabatic index affects critically the rate of gas accretion.

Assuming an environment where the heavy elements are homogeneously distributed, the envelope is unstable to convection when the Schwarzschild criterion
\begin{equation}
    \label{Schwarzschild}
    \nabla_{\rm rad}  > \nabla_{\rm ad}
\end{equation}
is satisfied, where we ignore compositional gradients in our analysis as it is found to drive convection on negligible scales and only at the very bottom of the envelope for mini-Neptunes \citep[see][]{Lee_2014}.\footnote{While \cite{Misener_2022} argue that the compositional gradient due to core envelope interactions can inhibit convection near the core, given the lack of mixture EOS at the $T$ and $P$ at the formation conditions, we defer these considerations to future research.} Thus we use $\nabla=$min$(\nabla_{\rm ad},\nabla_{\rm rad})$ in Equation (\ref{dT}). Throughout this work, we will refer to the boundary between the innermost convective zone and the overlying radiative region as the radiative-convective boundary (rcb). 

It has been shown that opacity has a sizable effect on planetary accretion rates \citep[e.g.,][]{Ikoma,Piso_2015}. We adopt the same opacity table that is used in \citet{Lee_2014} which is an expansion of the calculation of \cite{Ferguson} with the smoothing and extrapolation scheme outlined in \citet{Lee_2014}, their Section 2.1.3 for $\log T ({\rm K}) \geq 2.7$. Both dusty and dust-free models are considered where in the former, dust grains follow the interstellar medium (ISM) size distribution and in the latter, grains do not contribute to the opacity due to, e.g., coagulation and rain out \citep{Mordasini_2014,Ormel_2014}. Below $\log T ({\rm K}) = 2.7$ where the \citet{Ferguson} opacity table cuts off, we extrapolate following $\kappa \propto T^2$ for dusty and stitch the calculation of \citet{Freedman_2014} for dust-free \citep[see][their Section 2.2]{Lee_2015}. Over time, we may expect the disk gas to gradually transition from dusty to dust-free condition. Given the uncertain dust physics however, we opt for prudence and examine both dusty and dust-free scenarios.

\subsection{Boundary conditions}
\label{Boundary}

We describe the outer boundary of a planet with core mass $M_{\rm core}$ with the Hill radius

\begin{align}
\label{RH}
\begin{split}
    R_H&=\Big[\frac{(1+\mathrm{GCR})M_{\rm core}}{3M_\odot} \Big]^\frac{1}{3}a
    \\
    & \simeq 40R_\oplus \Big[\frac{(1+\mathrm{GCR})M_{\rm core}}{5M_\oplus}\Big]^{\frac{1}{3}}\big(\frac{a}{0.1 \mathrm{AU}}\big),
\end{split}
\end{align}
and the Bondi radius

\begin{align}
\label{Rb}
\begin{split}
    R_{B}&=\frac{G(1+\mathrm{GCR})M_{\rm core}}{c_s^2}
    \\
    &\simeq 90 R_{\oplus} \Big[ \frac{(1+\mathrm{GCR})M_{\rm core}}{5M_{\oplus}} \Big] \Big(\frac{\mu_d}{2.37} \Big) \Big(\frac{1000K}{T_d} \Big)
\end{split}
\end{align}
where $c_{\rm s} = \sqrt{kT_{\rm d}/\mu_{\rm d} m_{\rm H}}$ is the sound speed, $k$ is the Boltzmann constant, $T_{\rm d}$ is the disk midplane temperature, $\mu_{\rm d}$ is the mean molecular weight, and $m_{\rm H}$ is the mass of the hydrogen atom. The disk parameters $c_{\rm s}$, $T_{\rm d}$ and $\mu_{\rm d}$ are all evaluated at $a$.

We fix the outer boundary of the envelope  at the minimum of those two radii
\begin{equation}
   \label{adv_radius}
    R_{\rm adv}=\alpha_{\rm adv}R_{\rm out}= \alpha_{\rm adv}\min(R_H,R_B),
\end{equation}
with the inclusion of a free parameter $\alpha_{\rm adv} \leq 1$, which accounts for a flow-dominated region of the atmosphere ($R_{\rm adv}<r<R_{\rm out}$) where the outer disk injects entropy into the system. The thermal state of this flow-dominated region will be developed in Section \ref{advection_region}. As for the inner boundary down to which we integrate Equations (\ref{dM})-(\ref{dT}), we use a core radius which scales as \citep{VALENCIA2006545}

\begin{equation}
    R_{\rm core}=R_\oplus \Big(\frac{M_{\rm core}}{M_\oplus} \Big)^\frac{1}{4}.
    \label{R_core}
\end{equation}

The center of our planet is placed at the disk midplane for which the fiducial parameters $T_{\rm mid}$ and $\rho_{\rm mid}$  are taken from the minimum-mass extrasolar nebula (MMEN) of \cite{Chiang_Lau} modified for the irradiated disk profile of \cite{Chiang_Goldreich_1997}:

\begin{equation}
    \label{density_disk}
    \rho_{\rm mid}=6 \times 10^{-6} \Big( \frac{a}{0.1 \mbox{AU}} \Big)^{-2.9} \mbox{g/cm}^3
\end{equation}

\begin{equation}
    \label{temperature_disk}
    T_{\rm mid}=1000 \Big(\frac{a}{0.1 \mbox{AU}}\Big)^{-\frac{3}{7}} \mbox{K }.
\end{equation}
Since we are interested in revisiting the ability of advective flows in delaying the accretion timescale in gas-rich environment, we limit our calculations to the gas-full (i.e., not depleted) disk profiles. We assume that the disk midplane parameters are constant in time over the duration of gas accretion which not only simplifies the computations, but is also justified given the weak dependence of accretion rate on outer nebular conditions \citep[e.g.,][]{Lee_2014,Ginzburg_2016}, as long as the nebular gas density does not deplete by more than 8 orders of magnitude \citep{Lee_2018}. 
We verify a posteriori that this also applies in our model.

\subsection{Thermal state of the outer shells}
\label{advection_region}

To account for strong advective flows from the disk penetrating the planetary envelope, we include an outer advection region of the envelope ($R_{\rm adv}<r<R_{\rm out}$) dominated by entropy advection which we treat separately from the inner shells. The thermal state of this region of the envelope is determined by whether the cycling gas injected from the disk is allowed to mix with the surrounding fluid and cool down radiatively before being ejected back to the disk. We account for this entropy advection using mixing length theory \citep[e.g.,][]{Kippenhahn_2013}.

The total energy flux from both advection and radiation in this dynamical regime is given by the radiative gradient $\nabla_{\rm rad}$ required assuming fully radiative cooling:

\begin{equation}
    F_{\rm adv} + F_{\rm rad} = \frac{16 \sigma G}{3} \frac{T^4M_{\rm tot}}{\kappa P r^2} \nabla_{\rm rad}
    \label{Tot_E_flux}.
\end{equation}
Note that we use the total planetary mass $M_{\rm tot} = (1+ \mathrm{GCR})M_{\rm core}$ ignoring the small gas mass present above $R_{\rm adv}$ (this assumption is justified by the centrally concentrated mass distribution shown in Figure \ref{profiles_example}). The radiative energy flux,

\begin{equation}
    F_{\rm rad} = \frac{16 \sigma G}{3} \frac{T^4M_{\rm tot}}{\kappa P r^2} \nabla
    \label{F_rad}
\end{equation}
is given by the dimensionless local temperature gradient $\nabla$ (see Equation \ref{dT}).  Considering the advective mixing length $l_{\rm m}$ of the gas parcel, its average velocity in the medium and the work done over its motion, the local flux of advective energy is 

\begin{equation}
    \label{F_con}
    F_{\rm adv} = \rho v_{\rm adv} c_{\rm P} T  \frac{l_{\rm m}}{2H_{\rm P}} (\nabla - \nabla_{\rm p}).
\end{equation}
Here, $v_{\rm adv}$ is the velocity of the advective flow which we fix to the shear velocity of the gas at the outer boundary $v_{\rm adv} = R_{\rm out} \Omega$  with orbital angular frequency $\Omega$, $c_{\rm P} = \frac{P}{\rho T \nabla_{\rm ad}}$ is the specific heat per unit mass (at constant $P$) of the gas, $H_{\rm P} = \frac{P}{\rho g} $ is the pressure scale height with gravitational acceleration $g = \frac{G M_{\rm tot}}{r^2}$, and $\nabla_{\rm p} \equiv \Big( \frac{d \ln T}{d \ln P} \Big)_{\rm p}$ is the dimensionless gradient over the motion of the parcel. Since we assumed advective flows dominating the envelope down to $R_{\rm adv}$, we fix $l_{\rm m} \approx R_{\rm out} - R_{\rm adv}$.

Equations (\ref{Tot_E_flux}-\ref{F_con}) can be combined into a dimensionless form 

\begin{equation}
    \nabla + \frac{1}{U} (\nabla - \nabla_{\rm p}) = \nabla_{\rm rad},
    \label{Dimensionless_flux_equation}
\end{equation}
where we define a dimensionless parameter $U = \frac{32 \sigma T^3}{3 \kappa \rho^2 v_{\rm adv} c_P l_m}$. Following the derivation of \citet[][see their Section 7.1 and 7.2]{Kippenhahn_2013}, we have

\begin{equation}
    \nabla_{\rm p} - \nabla_{\rm ad} =  \frac{9}{4}U (\nabla - \nabla_{\rm p}),
    \label{Parcel_gradient}
\end{equation}
which allows Equation (\ref{Dimensionless_flux_equation}) to be solved for $\nabla$ in an explicit form

\begin{equation}
    \nabla = \frac{(9U^2+4U)\nabla_{\rm rad} + 4\nabla_{\rm ad}}{9U^2+4U+4}.
    \label{grad_outer}
\end{equation}

 \begin{figure}[t!]
\includegraphics[width=0.473\textwidth]{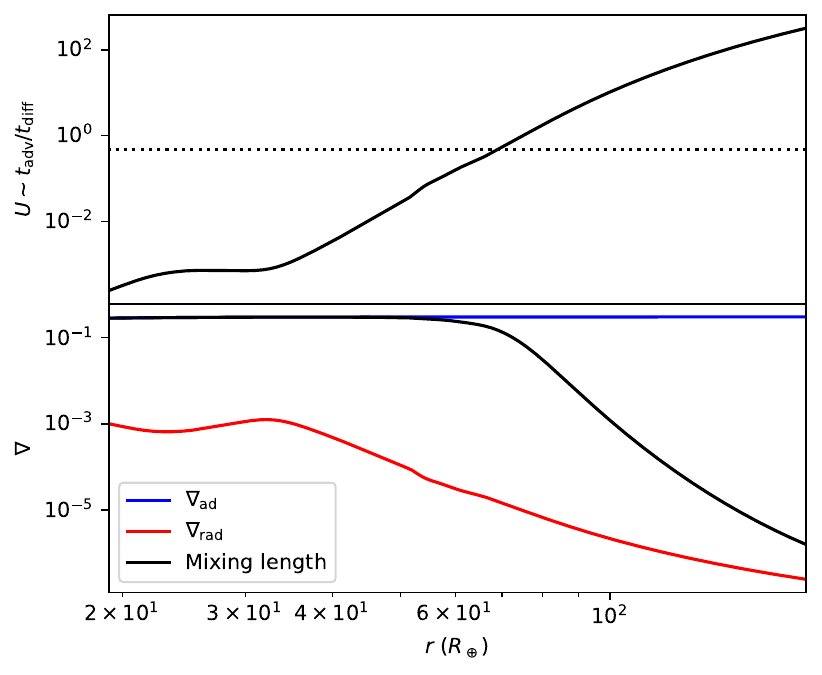}
 \caption{Transition between isentropic and radiative advection zones of the gas envelope  atop a $5M_\oplus$ core. In this case, a planet with a total gas-to-core mass ratio of 0.07 is embedded in a minimum mass extrasolar nebula with dust-free opacity at 1 AU. Upper panel: the dimensionless parameter $U$ which corresponds to the ratio of the transport timescale and the cooling time of advective flows as a function of radius ($r$) within the outer advective layer ($R_{\rm adv} < r < R_{\rm out}$). Lower panel : Dimensionless temperature gradient $\nabla \equiv d$ ln $T/d$ ln $P$ (black) as given by the mixing length based hybrid model of advection of Equation (\ref{grad_outer}) over the same region. The radiative diffusion gradient $\nabla_{\rm rad}$ (red) and the convection gradient $\nabla_{\rm ad}$ (blue) are shown as comparison. Transitions between radiative and isentropic cooling within the advective layers happen when $U \approx 0.5$ (dotted) which in this case corresponds to $r \approx 70 R_\oplus$.}
\label{U_figure}
\end{figure}

The value of parameter $U$ throughout the outer advective layers of the envelope  directly determines whether the advective flows will cool ($\nabla \rightarrow \nabla_{\rm rad}$) or not ($\nabla \rightarrow \nabla_{\rm ad}$), and can be linked to the transport timescale of the flow $t_{\rm adv} = \frac{l_m}{v_{\rm adv}}$ and its cooling timescale $t_{\rm diff} = \frac{E_{\rm diff}}{L}$. Here, 

\begin{equation}
    E_{\rm diff} = 4\pi r^2 l_m \rho c_{\rm P} (T_{\rm ad} - T)  = \frac{2\pi \rho c_P l_m^2 r^2 T}{H_{\rm P}} (\nabla - \nabla_{\rm p})
    \label{E_diff}
\end{equation}
 is the thermal energy that must be dissipated for the envelope temperature $T$ to be below the isentropic extension of the disk $T_{\rm ad} = T + \frac{l_m}{2} \frac{T}{H_{\rm P}}(\nabla - \nabla_{\rm p})$. With $L$ given by Equation (\ref{grad_rad}), we find

\begin{equation}
    t_{\rm diff} = \frac{3\kappa \rho^2 c_P l_m^2}{32\sigma T^3} \frac{(\nabla - \nabla_{\rm p})}{\nabla_{\rm rad}},
    \label{t_diff}
\end{equation}
so that
\begin{equation}
    \frac{t_{\rm adv}}{t_{\rm diff}} = U \frac{\nabla_{\rm rad}}{(\nabla - \nabla_{\rm p})} \sim U,
    \label{U_timescales}
\end{equation}
where we check a posteriori that $\frac{\nabla_{\rm rad}}{(\nabla - \nabla_{\rm p})} \sim 1$. Our approach is similar to the $\beta$ cooling model of \cite{Kurokawa} as the parameter $U$ considers the impact of radiative cooling on recycling flows. However, the main difference is that we do not artificially fix a value of $U$ ranging from an isentropic setup ($U\ll1$) to a fully radiative one ($U\gg1$). Instead, our model self-consistently solves for $U$ throughout the outer envelope.

Based on Equation (\ref{grad_outer}), transitions between isentropic advective layers and radiative ones happen when $9U^2 +4U \sim 4$ or $U \sim 0.5$. While we find that most envelopes throughout the parameter space explored in this work are subject to fully isentropic advection, Figure \ref{U_figure} shows how such transition happens in dust-free envelopes at 1 AU. In this case, advective flows that are strong enough to reach below $r \lesssim 70R_\oplus$ are cycled out without having time to cool while weaker flow of gas will thermally relax.

 Solving Equation (\ref{grad_outer}) for $\nabla (r)$ through the advective layer accounts for partial radiative cooling of the cycling gas and differs from the adiabatic treatment ($\gamma =1.4$) of this region of the envelope used by \cite{Ali_Dib_2020}. Taking $\nabla$ from Equation (\ref{grad_outer}) instead of the Schwarzschild criterion (Equation \ref{Schwarzschild}), we solve the structure equations over the outer envelope (from $R_{\rm out}$ to $R_{\rm adv}$) with the mass fixed to $M_{\rm tot}$ (Equations \ref{dP}-\ref{dT}).

\subsection{Connecting snapshots in time}
\label{snapshot}

Like \citet{Piso}, we take the planet luminosity to be spatially constant which implicitly assumes the luminosity generated in the outer region to be minimal, which we verify a posteriori, even when we account for entropy advection. For a fixed GCR, there is a constant luminosity eigenvalue $L$ for the stellar structure equations (Equations \ref{dM}-\ref{dT}). We integrate these equations from $R_{\rm adv}$ to $R_{\rm core}$ with the boundary conditions at $R_{\rm adv}$ described by the thermal state of the advective layer (Section \ref{advection_region}). We iteratively solve for $L$ until the mass profile agrees with the fixed GCR value within a relative difference of 1\% for each specific snapshot.

We connect the resulting snapshots together by computing the time it takes to cool from one snapshot to next. Although the planetesimal accretion of solids forming the core can act as an battery, the range of planetesimal accretion rate that can successfully avoid runaway is extremely limited over 0.1--5 AU and the typical solid accretion rate in solar nebula would in fact accelerate the runaway by way of growing the core to too high a mass \citep[see][their Figure 2]{Lee_2015}. Furthermore, at $\lesssim$1 AU, the core coagulation timescale is shorter than the envelope cooling time \citep[see, e.g.,][their equation 2]{Lee_2014} and so we ignore the thermal energy in the core. Changes to the energy budget of the envelope therefore come directly from its cooling and accretion of gas. Since most of the envelope mass is centrally concentrated (owing to H$_2$ dissociation driving adiabatic index $\gamma < 4/3$) and the surface lid of the convective zone acts as a thermal bottleneck, we can characterize the thermal state of the envelope with the state variables at the radiative-convective boundary $R_{\rm rcb}$. We follow the cooling treatment of \citet{Piso} for the change in energy which gives the time elapsed between snapshots:

\begin{equation}
    \label{delta_t} 
    \Delta t = \frac{-\Delta E + \langle e_M \rangle \Delta M- \langle P \rangle \Delta V_{\langle M\rangle}}{\langle L \rangle}.
\end{equation}
Here, the average and difference of a quantity Q over two snapshots are denoted with $\langle Q \rangle$ and $\Delta Q$ respectively, and $E$ is the total energy of the envelope 

\begin{equation}
\label{Energy} 
    E = -\int\frac{GM(<r)}{r}dM + \int U dM
\end{equation}
where we account for the specific internal energy ($U$) contributions with our EOS. The integral is carried out from $R_{\rm core}$ to $R_{\rm rcb}$ to cover the entire inner convective zone. Energy from mass accretion is given by

\begin{equation}
\label{e_M} 
    e_M = -\frac{GM}{r}\Big|_{R_{\rm rcb}} +  U \mid_{R_{\rm rcb}}.
\end{equation}

The last term in the numerator of Equation (\ref{delta_t}) accounts for the change in the volume $V_{\langle M\rangle}$ of the innermost convective zone with the pressure taken along its surface. We do not ignore the contribution of the surface energy terms $\langle e_M \rangle \Delta M$ and $\langle P \rangle \Delta V_{\langle M\rangle}$ as excluding these terms can underestimate the GCR by factors of $\sim$10\% and artificially delay the runaway. We discuss this effect more in Section \ref{Discussion}.

Since we only compute time intervals, we need to fix an initial time $t_0$ which we set to the Kelvin-Helmholtz time of the first snapshot $|E|/L$ which is small ($\lesssim 0.05$ Myr) compared to the disk lifetime of $\sim 10$ Myr \citep{Mamajek_2009,Michel_2021}. 

\section{Results}
\label{results}

\subsection{Effect of advection on envelope structure and cooling}
\label{Hybrid_results}

\begin{figure}[t!]
\includegraphics[width=0.473\textwidth]{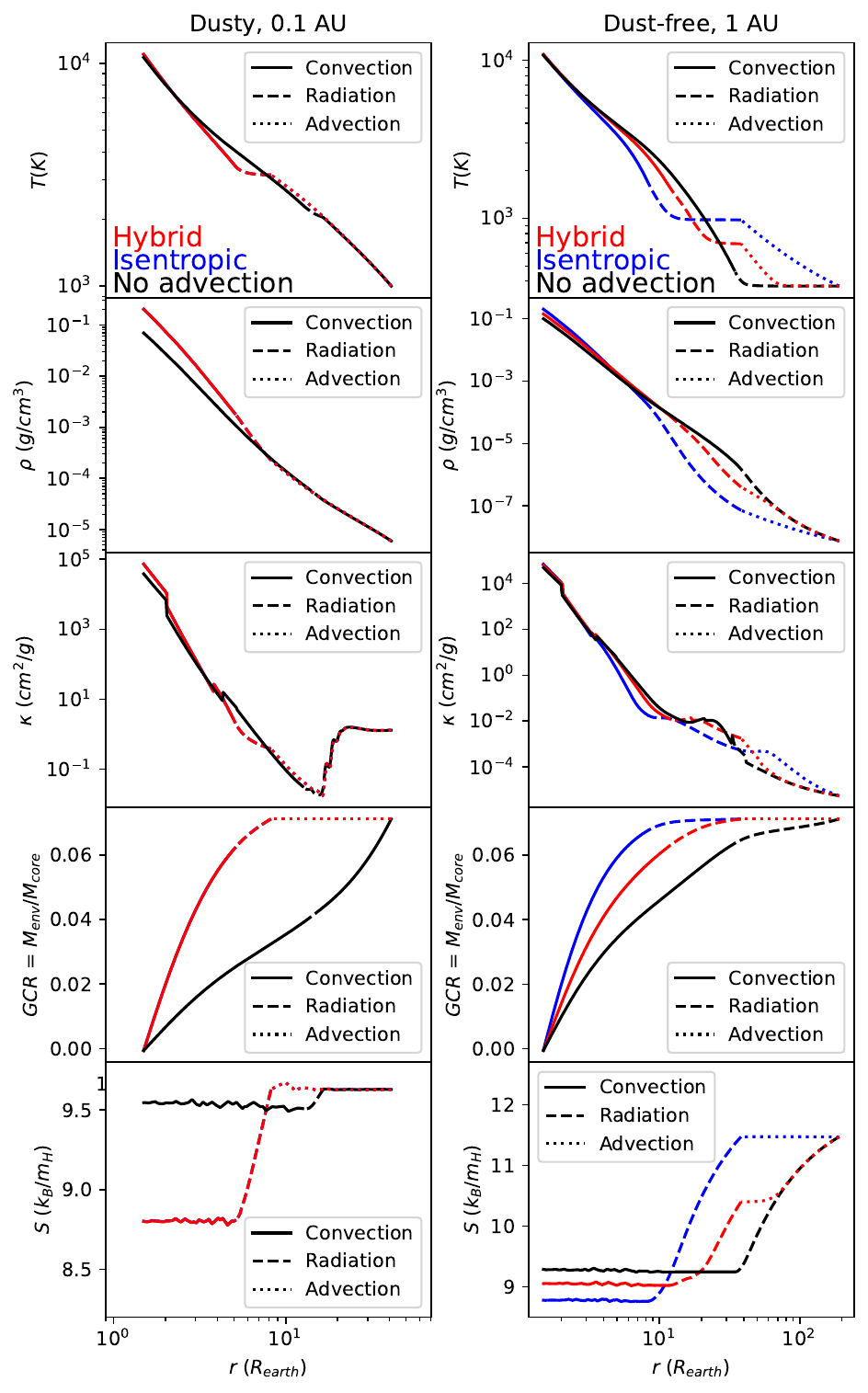}
 \caption{Radial profiles of the state variables of the gas envelope atop a $5M_\oplus$ core embedded in a minimum mass extrasolar nebula with a total gas-to-core mass ratio (GCR) of 0.07. The left and right panels show respectively systems at 0.1 AU with dusty opacity and 1 AU with dust-free opacity. From top to bottom, the temperature, density, opacity, GCR and entropy radial profiles are displayed for envelopes with hybrid (red), isentropic (blue) and without (black) entropy advection from the disk. We note that the small entropy variations in isentropic layers are due to numerical noise in the EOS grid. The advection ratio ($\alpha_{\rm adv}=R_{\rm adv}/R_{\rm out}$) is fixed to 0.2. Dotted curves indicate the outer region dominated by advective flows while radiation and convection dominated zones of the inner envelope are represented by dashed and solid curves respectively. Under the hybrid treatment, the outer advective region is fully isentropic with the disk for dusty envelopes at 0.1 AU, and it is almost fully radiative for dust-free envelopes at 1 AU.}
\label{profiles_example}
\end{figure}

With the one-dimensional differential system developed in Sections \ref{equation_system} and \ref{Boundary} and the hybrid characterization of the advective flows penetrating the envelope explored in Section \ref{advection_region}, we can solve for snapshots of the thermal state of the envelope for any core mass, orbital distance, different opacity regimes, advection depth $\alpha_{\rm adv}$ and total GCR. Figure \ref{profiles_example} shows the resulting temperature, density, opacity, entropy and envelope mass profiles under this hybrid scheme for a $5M_\oplus$ core embedded in a gas-rich nebula at both 0.1 AU with dusty opacities and 1 AU with dust-free opacities. While we explore the full parameter space, we choose these two limiting cases for illustration purpose: the former envelope is characterized by a fully isentropic entropy advection region while the latter envelope is allowed to mostly cool by radiation according to our hybrid treatment of the outer advective layer (Section \ref{advection_region}).

\begin{figure}[t!]
\includegraphics[width=.473\textwidth]{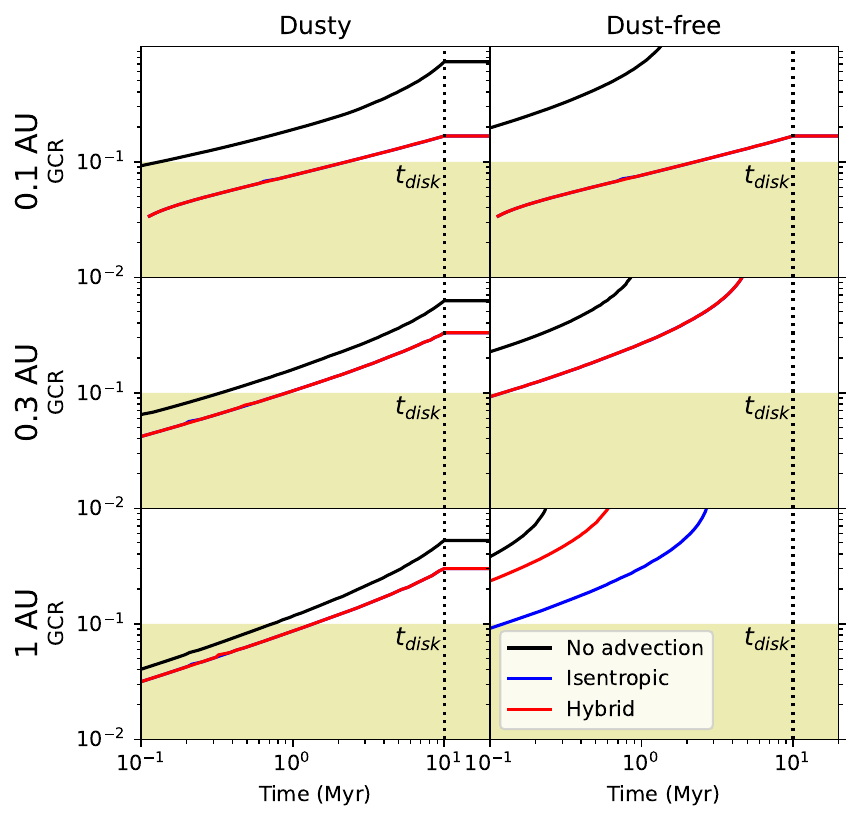}
\caption{Time evolution of the gas-to-core mass ratio (GCR $\equiv M_{\rm gas}/M_{\rm core}$) of a $5M_\oplus$ core embedded in a minimum mass extrasolar nebula (MMEN) for an advection ratio of $\alpha_{\rm adv}=0.3$. Left and right panels describe models with dusty and dust-free opacities respectively. Upper panels show the evolution of a core at orbital distance of 0.1 AU, middle panels 0.3 AU and lower panels 1 AU.  The model with a fully isentropic (blue) outer region is compared to the hybrid scheme (red). The evolution of the envelope mass of models without recycling (black), the target mass range for mini-Neptunes envelope mass (shaded region) and the protoplanetary disk lifetime (vertical dotted line) are indicated as references. Note that the hybrid advection model overlaps with the fully isentropic case in all cases except for dust-free opacity at 1 AU. Overall, we find that entropy advection alone is unable to reproduce mini-Neptunes.} 
\label{hybrid_vs_radiative_vs_isentropic}
\end{figure}

In both 0.1 AU and 1 AU cases, we find that accounting for the advective flows result in steeper envelope profiles with more centrally concentrated mass and smaller internal specific entropy. These steep profiles emerge because for a given envelope mass, an outer penetrative flow shrinks the spatial volume within which the gas mass can be packed. While changes to the temperature at the rcb are small between models with and without outer advective zones, the density at the rcb rises in the former which increases the optical depth and so we expect a delay in cooling time when advection is taken into account. Compared to 0.1 AU, we see that the changes to the thermal structures at 1 AU by advective flows are more muted. At wider orbits, the advective flows can cool easily and appear nearly isothermal, similar to the upper envelopes of non-advective, unrestricted cases. For the same core mass and gas mass, the Hill sphere is larger at 1 AU; there is more room to pack the same amount of gas mass and so the envelope profiles are less affected by the changes in the outer envelope. We see from Figure \ref{profiles_example} that forcing an isentropic profile at 1 AU would create an even steeper inner envelope structure overestimating the density at the rcb.

Figure \ref{hybrid_vs_radiative_vs_isentropic} illustrates the degree of delay in gas accretion by cooling caused by the entropy advection. For $a \geq 0.3 \mathrm{AU}$,  the effect is minor, either causing factors of $\lesssim$2 changes in the final GCR for dusty accretion or being unable to avoid the runaway gas accretion as compared to the unrestricted evolution in dust-free cases. As mentioned previously, the smaller Hill sphere at shorter orbital distances forces a significantly steeper envelope structure when advective flows are taken into account and the higher rcb density leads to higher optical depth and so slower cooling. Even so, we find that a 5$M_\oplus$ core that begins accreting gas in early, gas-rich environment will always (even for dust-free accretion at 0.1 AU where runaway accretion is successfully prevented) end up with GCR $\gtrsim$0.15 (for our fiducial $\alpha_{\rm adv} = 0.3$) which is larger than the expected envelope mass fraction of mini-Neptunes. If dust grains do not contribute to opacity during accretion, such a core is expected to undergo runaway gas accretion and blow up into gas giants even at 0.3 AU, even with the recycling flows. We therefore conclude that entropy advection alone is insufficient in limiting the rapid gas accretion onto high-mass mini-Neptune cores for $\alpha_{\rm adv} \gtrsim 0.3$. The feasibility and realism of preventing runaway accretion with smaller $\alpha_{\rm adv}$ is explored in Sections \ref{Scalings} and \ref{realistic_alpha_adv}. Our use of realistic EOS, opacity, and cooling time calculation arrive at a result that differs from the previous study by \citet{Ali_Dib_2020}, which we discuss in more detail in Section \ref{effect_kappa_eos}.

\subsection{Scaling relations for GCR($t$,$\alpha_{adv}$)}
\label{Scalings}

\begin{figure*}
\subfigure[Dusty]{\includegraphics[width=9cm]{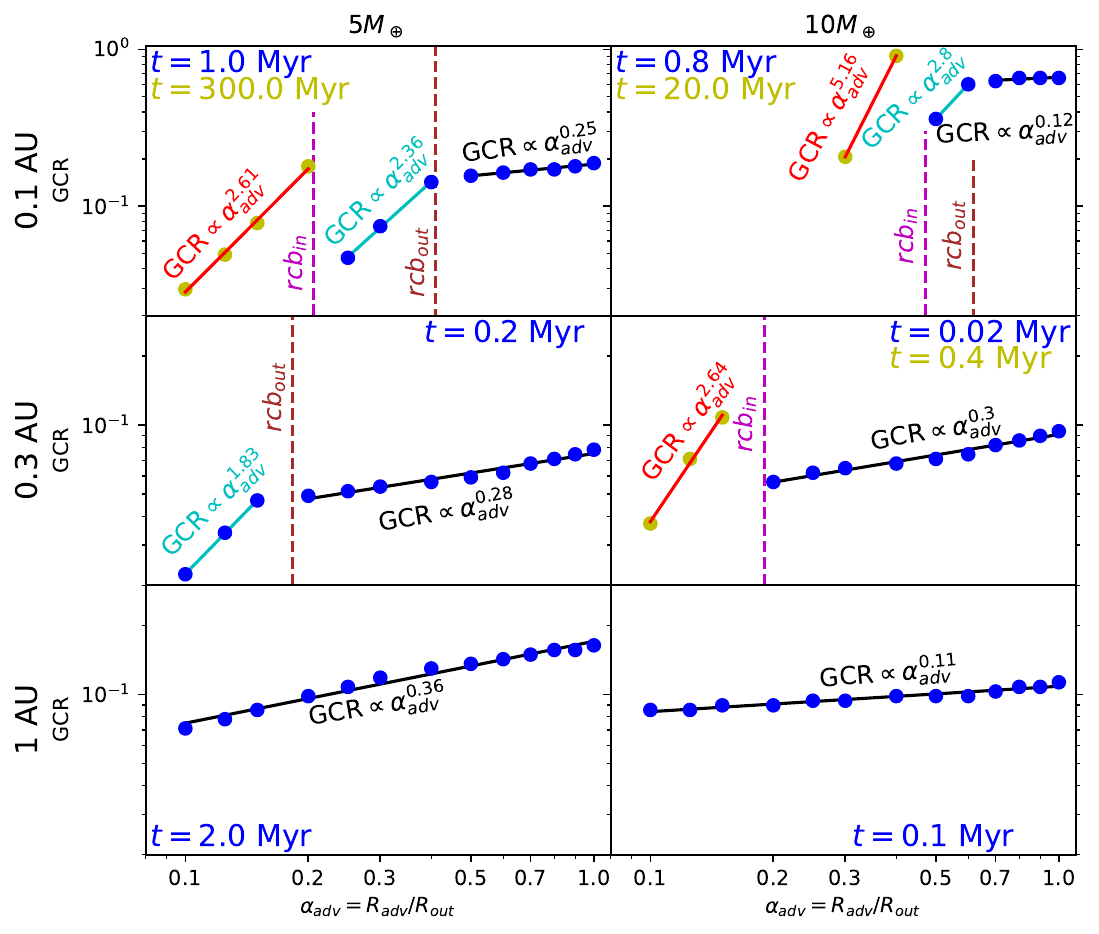}}
\subfigure[Dust-free]{\includegraphics[width=9cm]{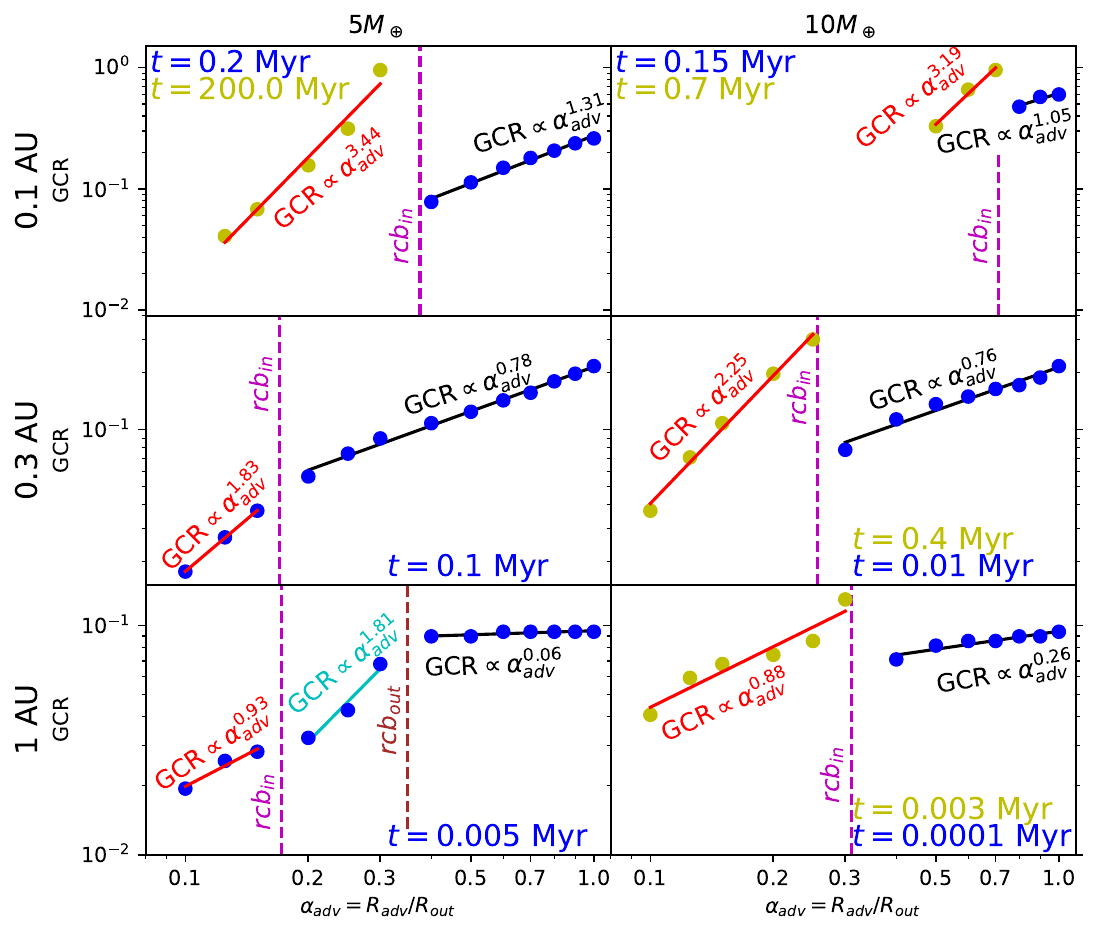}}

\caption{Gas-to-core mass ratio (GCR) of a core embedded in a gas-rich nebula at a fixed time as a function of advection depth ($\alpha_{\rm adv}$) under different boundary conditions. The left and right figures account for dusty and dust-free opacities respectively. From top to bottom, the orbital distance is varied from 0.1 AU to 0.3 AU and 1 AU. The first column of each subfigure considers a $5M_\oplus$ core while the second one a $10M_\oplus$ core. For each setup, the fixed time at which the GCR's are taken is annotated in blue text and those GCR's indicated with blue circles. In cases for which the GCR could not be obtained at a same instant for all values of $\alpha_{\rm adv}$ (refer to Figure \ref{GCR_vs_t_all}), such GCR's are evaluated at a later time (yellow text) and indicated with yellow circles. In general, the GCR-$\alpha_{\rm adv}$ scaling changes its behavior when $\alpha_{\rm adv}$ becomes smaller than the locations of rcb's in the original unrestricted envelopes, which are denoted with magenta (rcb$_{\rm in}$) and brown (rcb$_{\rm out}$) vertical lines. Empirically-fit scalings $\mathrm{GCR} \propto \alpha_{\rm adv}^\theta$ are annotated in each ranges of $\alpha_{\rm adv}$.}

\label{GCR_vs_alpha_adv}
\end{figure*}

With the realistic treatment of atmospheric recycling developed in this work, we update the previously developed semi-analytic scaling relationships between GCR, time, and the penetration depth of advective flows. In unrestricted systems, \cite{Lee_2015} derived a general expression for the evolution of the atmospheric mass in terms of state variables and fiducial parameters of the model. The goal of this section is to empirically obtain a scaling of form 

\begin{equation}
    \label{GCR_prop_to}
    \mathrm{GCR} \propto t^\frac{1}{2+\alpha} \alpha_{adv}^\theta,
\end{equation}
where $\kappa_{\rm rcb} \propto \rho_{\rm rcb}^\alpha$ and $\alpha \sim$ 0.15-0.7 over the parameter space we explore. Because we fix the core mass, we do not refit the scaling on $M_{\rm core}$ (but we check a posteriori that the GCR-$M_{\rm core}$ scaling derived by \citealt{Lee_2015} matches well our calculations between 5$M_\oplus$ and 10$M_\oplus$, shown in Figure \ref{GCR_vs_t_all} in the Appendix). Other dependencies such as the adiabatic gradient and mean molecular weight are not directly discussed here as they do not vary significantly across our simulations; we also do not discuss the effect of metallicity in this paper as we have fixed our calculations to solar metallicity. The GCR-$T_{\rm rcb}$ scaling (where $T_{\rm rcb}$ is the envelope temperature at $R_{\rm rcb}$) depends on $\kappa-T$ scaling which we find to vary non-trivially as we vary $\alpha_{\rm adv}$ since the $T_{\rm rcb}$ ranges from $\sim$2000--3000 K where the $\kappa_{\rm rcb}-T$ scaling probes the transition to H- opacity (i.e., from a weak dependence to a near exponential). We therefore subsume the dependence on $T_{\rm rcb}$ under GCR-$\alpha_{\rm adv}$ scaling which we fit against our numerical results.

We empirically fit for $\theta$ over numerical results computed at $\alpha_{\rm adv} \in [$0.1, 0.125, 0.15, 0.2, 0.25, 0.3, 0.4, 0.5, 0.6, 0.7, 0.8, 0.9, 1$]$ at a fixed time where all GCR values of $\alpha_{\rm adv}$ are pre-runaway. The resulting values of GCR as a function of advection depth are presented in Figure \ref{GCR_vs_alpha_adv}. In some instances at 0.1 AU where the delays due to advection are significant (usually for $\alpha_{\rm adv} \lesssim 0.3$), the GCR for low $\alpha_{\rm adv}$ had to be taken at a later time (as otherwise the envelope is fully convective). We observe that the GCR as a function of $\alpha_{\rm adv}$ behaves differently for different advection depths, whereby the effect of $\alpha_{\rm adv}$ becomes more significant when the advective flows penetrate below the rcb's in the original unrestricted envelopes, similar to what was reported by \citet{Lee_2022}.

The calculation of \citet{Lee_2022} differs from ours in two ways. First, their focus was on gas-depleted environment. Second, they accounted for advective flows by simply shrinking $R_{\rm out}$. To more directly compare our results with \citet{Lee_2022}, we reduce the MMEN disk density by a factor of 0.01. At 0.3 AU in dust-free atmospheres, we find that entropy advection has no effect on the envelope for  $\alpha_{\rm adv} \gtrsim 0.3$ as the outer advective layer is isothermal and identical to the case of $\alpha_{\rm adv} =1$ with all else equal. For $\alpha_{\rm adv} \lesssim 0.3$, we find a dependency on advection depth of GCR $\propto \alpha_{\rm adv}^{1.3 \pm 0.1}$ . The dusty analogues behave differently with a fully isentropic advective layer and an overall scaling of $\propto \alpha_{\rm adv}^{0.49 \pm 0.03}$. These behaviours differ from the scaling on outer boundary of GCR $\propto R_{\rm out}^{0.27 \pm 0.05}$ and $\propto R_{\rm out}^{0.31 \pm 0.08}$ over the whole envelope for dusty and dust-free envelopes respectively of \citet[][see their Figures 4 and 5]{Lee_2022}. These results indicate that simply changing the $R_{\rm out}$ is not equivalent to a more careful treatment of the advective flows.

In gas-rich environments (which is our focus), our unrestricted dusty envelopes are characterized by outermost and innermost convective zones with a radiative window sandwiched in-between, so that there are two rcb's: the outer one (rcb$_{\rm out}$) where the envelope transitions from a convective to a radiative zone outside-in; and the inner one (rcb$_{\rm in}$) where the envelope transitions from a radiative to a convective zone outside-in. As demonstrated in Figure \ref{GCR_vs_alpha_adv}, we see that $\theta$ is the largest (i.e., GCR is affected most strongly) when the flow can penetrate inside rcb$_{\rm in}$  of the corresponding $\alpha_{\rm adv}=1.0$ envelope with all else equal. This behavior is expected because the envelope is forced to be confined within a radius that is smaller than the initial rcb$_{\rm in}$, significantly increasing the density (and also the temperature) at the new rcb, delaying the cooling process. When the flow penetrates below the original ($\alpha_{\rm adv} = 1.0$) rcb$_{\rm out}$ but above the original rcb$_{\rm in}$, the GCR-$\alpha_{\rm adv}$ becomes weaker, and when the flow cannot penetrate even the outer rcb, the GCR-$\alpha_{\rm adv}$ scaling becomes even weaker.

 \begin{figure}[t!]
 \centering
\includegraphics[width=0.473\textwidth]{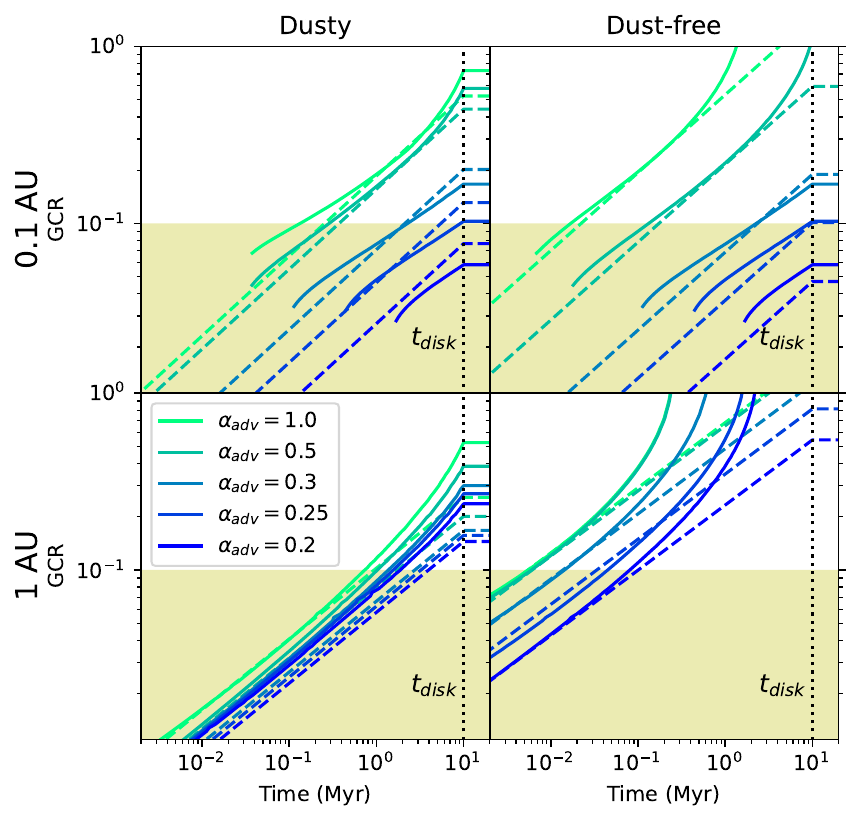}
\caption{Time evolution of the gas-to-core mass ratio (GCR) of a 5 $M_\oplus$ core embedded in a gas-rich nebula (solid lines) compared to the semi-empirical scalings (dashed lines) of Equation (\ref{GCR_prop_to}) with the appropriate $\theta$ from Figure \ref{GCR_vs_alpha_adv}.  The advection depth ($\alpha_{\rm adv}$) is varied from 1 to 0.2 as indicated by the color scheme. The target range for mini-Neptune envelope mass fraction (shaded region) and the protoplanetary disk lifetime ($t_{\rm disk}$, dotted vertical line) are indicated as references. We consider accretion to have completed beyond $t_{\rm disk}$ and so beyond this point, GCR plateaus. In general, the scalings of Figure \ref{GCR_vs_alpha_adv} agree with the computed accretion rates within factors of $\sim20\%$ whereby the slight deviations arise from varying $\alpha$ for different $\alpha_{\rm adv}$. A full exploration of the parameter space over core mass, orbital distance, dusty vs.~dust-free opacity, and $\alpha_{\rm adv}$ is shown in Figure \ref{GCR_vs_t_all}.}
\label{scaling_examples}
\end{figure}

We demonstrate in Figure \ref{scaling_examples} that Equation (\ref{GCR_prop_to}) with the $\theta$ computed in Figure \ref{GCR_vs_alpha_adv} provides a good approximation to our numerical results. It is however not a perfect depiction of the effect of entropy advection as can be observed in the slightly decreasing slope of the GCR-t curves at 0.1 AU which implies an underlying dependency of the opacity-density scaling at the rcb ($\kappa_{\rm rcb} \propto \rho_{\rm rcb}^\alpha$) on advection depth. While the strong effect of entropy advection when the flow penetrates below the inner rcb could theoretically keep the final envelope mass small enough to be consistent with the observed properties of mini-Neptunes at 0.1 AU (see Figures \ref{scaling_examples} and \ref{GCR_vs_t_all}), it would require advection depths of $\alpha_{\rm adv} \lesssim 0.25$.

\section{Discussion}
\label{Discussion}

\subsection{Can advection alone produce mini-Neptunes?}
\label{realistic_alpha_adv}

The results presented in Section \ref{results} establish a minimal depth within the envelope at 0.1 AU for a $5 M_\oplus$ core down to which the outer entropy advection region must extend in order to produce mini-Neptune on its own (i.e. without being combined with other scenarios to delay atmospheric growth which we discuss in Section \ref{other_scenarios}). We must therefore refer to 3-D hydrodynamic models of advective flows to determine whether $\alpha_{\rm adv} \lesssim 0.25$ is a realistic advection depth in our parameter space.

In their 3-D radiative hydrosimulations, \cite{Lambrechts} consider advective flows about a $5 M_\oplus$ core with fixed opacities of $\kappa = 0.01 \, \mathrm{cm}^2\,\mathrm{g}^{-1}$ and $\kappa = 1 \, \mathrm{cm}^2\,\mathrm{g}^{-1}$, which are comparable to the ranges of $\kappa$ that we see at the rcb's (see Figure \ref{profiles_example}). In their low opacity regime, they report a three-layer envelope structure (see their Figures 10 and 12): an innermost convection layer ($r \lesssim 0.1 R_H$), a bound radiative shell ($0.1 R_H \lesssim r \lesssim 0.4 R_H$) and an outer unbound region dominated by advective flows ($0.4 R_H \lesssim r$). Noting that  $R_H = R_{\rm out}$ for a $5 M_\oplus$ core at 0.1 AU, such advective flows would not satisfy our criterion to limit the final GCR to less than 10\%. However, when opacity is increased to $1 \, \mathrm{cm}^2\,\mathrm{g}^{-1}$, \cite{Lambrechts} only observe this three-layer structure if there is no extra heating source (e.g., solid accretion). Otherwise, advective flows dominate the entire envelope of planets allowing for values of $\alpha_{\rm adv}$ well below 25 \%. While seemingly counterintuitive, they find that higher opacities improve radiative transport as the envelope puffs up, lowering central densities. However, solid accretion can also increase the rate of gas accretion by way of increasing the core mass (see Section \ref{snapshot} for relevant discussion), so we conclude that a depth of $\alpha_{adv} \lesssim 0.25$ would not be a realistic characterization of advective flows unless there is a fine-tuned extra energy source.

On the other hand, \cite{Moldenhauer_2021} and \cite{Moldenhauer_2022} argue that recycling flows could dominate the entire envelope thus fully allowing the possibility of $\alpha_{\rm adv} \lesssim 0.25$. To reduce their cooling timescales compared to their simulation time, they adopt $\kappa \lesssim 10^{-3} \mathrm{cm}^2\,\mathrm{g}^{-1}$ which is several orders of magnitude lower than the opacities we observe in our simulations (refer to Figure \ref{profiles_example}). Nonetheless, showing that the recycling timescale of the flow is independent of its cooling timescale, \cite{Moldenhauer_2022} argue that advection should dominate the entire envelope even with more realistic higher opacities.
 
 In their 3-D radiation-hydrodynamics models of protoplanet envelopes,  \cite{Bailey_2023} report advective flows dominating envelopes down to $r \sim 0.2-0.3 H_0$ (lower panels of their Figure 2)  where $H_0$ is the disk scale height ($H_0 \sim 1.16 R_H = 1.16 R_{\rm out}$ for a $5 M_\oplus$ core at 0.1 AU). Based on this result we cannot fully exclude the possibility of creating sub-Neptunes purely by advective flows at 0.1 AU as it is marginally do-able at $\alpha_{\rm adv} \sim 0.25$. However, the opacity used by \cite{Bailey_2023} is significantly lower ($\kappa \lesssim 6\times 10^{-4} \mathrm{cm}^2\mathrm{g}^{-1}$) than what we observe ($\kappa \gtrsim 10^{-2} \mathrm{cm}^2\mathrm{g}^{-1}$) in our envelope profiles which again makes it hard to draw direct comparison with their work.\footnote{ We convert the dimensionless opacity ($\kappa =100$) of \cite{Bailey_2023} to cgs units using our disk parameters $H_0$ and $\rho_0$ (see their Equation 4).}

It becomes apparent that direct comparison between 1-D semi-analytical thermal calculations and 3-D hydrodynamic simulations can be challenging. For better apples-to-apples comparison, one would require 3-D simulations of advective flows for low-mass planets at 0.1 AU using realistic opacities that vary with temperature and pressure \citep[e.g.,][]{zhu_2021,Zhu_2022}. While \cite{zhu_2021} do expect advection to have a larger impact at 0.1 AU, their focus is on large orbital distances $a > 5 \mathrm{AU}$. The challenge of hydrodynamic models in resolving the inner envelope where we find centrally concentrated mass and high opacities (c.f Figure \ref{profiles_example}) may also limit the accuracy of these 3-D simulations. At present, we cannot rule out the possibility that entropy advection, on its own, may produce mini-Neptunes for a 5 $M_\oplus$ core at 0.1 AU but emphasize the necessary condition of having $\alpha_{\rm adv} \sim 0.25$. At any larger distances (even at 0.3 AU), entropy advection alone will not be able to generate a mini-Neptune for any value of $\alpha_{\rm adv}$ over $\sim$0.1 (see Figure \ref{GCR_vs_t_all}).

\subsection{Importance of opacity and equation of state}
\label{effect_kappa_eos}

\begin{figure}[t!]
\includegraphics[width=.473\textwidth]{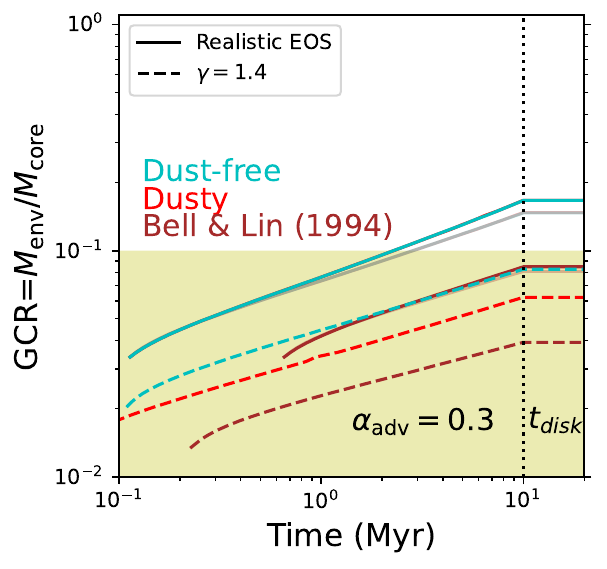}
\caption{Effect of opacity and equation of state (EOS) on the time evolution of the gas-to-core mass ratio (GCR) of a $5M_\oplus$ core embedded in a gas-rich nebula at 0.1 AU for $\alpha_{\rm adv} = 0.3$. The dusty opacity model (red) is compared to the dust-free one (cyan) and to the power-law opacity of \cite{Bell_Lin_1994} (brown).
Models using the simplified EOS of \cite{Ali_Dib_2020} with fixed adiabatic index $\gamma = 1.4$ and the realistic EOS of \cite{Lee_2014} are described by dashed and solid curves respectively. We show the effect of not accounting for the surface energy contributions in Equation (\ref{delta_t}) with lighter colors. Note that the light cyan and light red curves coincide and thus appear as gray.
The target GCR range for mini-Neptunes (shaded region) and the protoplanetary disk lifetime (dotted vertical line) are indicated as references. Simplifications in opacities, EOS, and the exclusion of surface energy contributions overestimate the impact of advection on accretion rates by a factor of $\sim$5.}
\label{op_eos}
\end{figure}

\begin{figure}[t!]
\includegraphics[width=9cm]{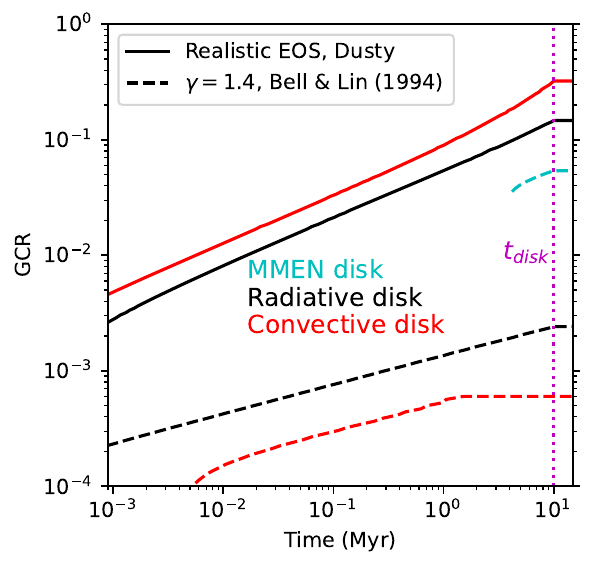}
\caption{Time evolution of the gas-to-core mass ratio (GCR) of a $10M_\oplus$ core at 0.1 AU with $\alpha_{\rm adv}=0.3$  under different disk conditions. The MMEN ($\rho_{\rm mid}=6 \times 10^{-6}$ g/cm$^3$, $T_{\rm mid}=1000$ K) disk used in this work is shown in cyan along with both the radiative ($\rho_{\rm mid}=7.45 \times 10^{-9}$ g/cm$^3$, $T_{\rm mid}=2000$ K) and convective ($\rho_{\rm mid}=4.96 \times 10^{-10}$ g/cm$^3$, $T_{\rm mid}=2000$ K) disks used by \cite{Ali_Dib_2020} in black and red respectively. Our model with realistic EOS and dusty opacity (solid) is compared the fixed adiabatic index of $\gamma =1.4$ and the power-laws opacity of \cite{Bell_Lin_1994} (dashed) used by \cite{Ali_Dib_2020}. The disk lifetime is indicated by a dotted vertical line. Under MMEN disk and our EOS/opacity, 10$M_\oplus$ cores' envelopes remain fully convective and so we cannot evolve them (see also Figure \ref{GCR_vs_t_all}. With realistic EOS and opacity, we do not observe any premature halting of gas accretion in contrast to what was reported by \citet{Ali_Dib_2020}.}
\label{disk_comparison}
\end{figure}

Our results highlight the importance of realistic EOS, opacities, and thermal state of recycling flows in quantifying the latter's effect on planetary envelope formation. Figure \ref{op_eos} demonstrates how simplifications of these elements can lead to an underestimation of gas cooling and therefore accretion rates. We replace the EOS of \cite{Lee_2014} used in the previous sections with the one of \cite{Ali_Dib_2020} which adopts a fixed adiabatic index of $\gamma=1.4$ and a mean molecular weight of $\mu = 2.34$. Doing so reduces the final values of GCR by a factor of $\sim$2.5-3, well and artificially within the mini-Neptune mass range. With our more realistic EOS, the adiabatic index is allowed to drop down to $\approx 1.2$ in the deep envelope as the energy is spent in dissociating H$_2$ molecules. This smaller adiabatic index establishes a more centrally concentrated mass and energy profile, allowing convection to prevail through a larger radial extent, ultimately reducing the pressure/density and therefore the optical depth at the rcb and consequently hastening envelope cooling. 

We investigate the effect of using realistic opacities by comparing the dusty and dust-free tables of \cite{Ferguson} to the power-law opacity of \cite{Bell_Lin_1994} with grains opacity suppressed by a factor of 10 as done by \cite{Ali_Dib_2020}. As illustrated in Figure \ref{op_eos}, adopting the more realistic opacities result in the final GCR larger by a factor $\sim$2 compared to assuming the power-law opacity of \citet{Bell_Lin_1994}. In both our opacity model and the one of \cite{Bell_Lin_1994}, the rcb appears inside the dust sublimation zone. The disparity between the two schemes is due to the more precise quantum treatment of molecular opacity in the calculations of \citet{Ferguson} which gives a stronger temperature-dependence of $\kappa$ ultimately decreasing $\kappa$ and therefore the optical depth of the rcb, again leading to faster cooling. We also note that the dusty and dust-free models are identical to each other at 0.1 AU as the outer advective region ($0.3 R_{\rm out}< r< R_{\rm out}$) in each opacity scheme is fully isentropic, and the inner envelope has temperatures above the H$_2$ dissociation threshold of 2500 K so that dust grain opacities are almost irrelevant. Replacing both our EOS and opacity by the ones of \cite{Ali_Dib_2020} limits the final GCR to $\sim 0.04$ which is $\sim$5 times lower than what we find with our realistic treatment of advection ($\mathrm{GCR} \sim$ 0.2).

Ignoring the surface energy sources in cooling timescale calculations also plays a role in overestimating the effect of entropy advection in limiting gas accretion. In particular, the energy gained from accreting gas between snapshots ($\langle e_M \rangle \Delta M$ term in Equation (\ref{delta_t})) increases when including advection, even before runaway accretion begins. This negative extra energy contribution can increase from $\lesssim$10\% of the total changes in the energy budget $\Delta E$ for $\alpha_{\rm adv}=1$ to $\sim$30 \% for $\alpha_{\rm adv}=0.3$. As shown in Figure \ref{op_eos}, including this extra energy source in cooling time calculations hastens accretion rates, with the final GCR higher by $\sim$ 10\%. While this variation is not as drastic as the impact of using realistic opacity and EOS, we still recommend including these surfaces energy sources to more accurately portray the thermal relaxation of the envelope.

\citet{Ali_Dib_2020} found that the density and the temperature of the disk can affect critically the gas accretion rate. For example, they report that within their convective disks, planets would quickly establish an inner isothermal region thereby completely halting accretion. In Figure \ref{disk_comparison}, we illustrate the effect of adopting different disk conditions on the rate of gas accretion, using the radiative and convective disks defined by \citet{Ali_Dib_2020}. Even with advection, accretion is never halted before the protoplanetary disk dissipates unlike the isothermal state reached by \cite{Ali_Dib_2020} after $\sim$5 Myr in the radiative disk and over the entire life of the disk in the convective disk (refer to their Figure 5 for comparison). We attribute the cause of this different result to our use of more realistic EOS, opacity and proper accounting of surface energy in the calculation of cooling timescale. Nevertheless, compared to the gas-full MMEN disk, there is a reduction in the final GCR in the radiative and convective disks, which are functionally equivalent to depleting the MMEN disks by factors of $\sim$800 and $\sim$1200 respectively \citep[see also][for self-consistently generating convective disks through magnetorotational instability]{Jankovic21,Jankovic_2022}. Such gas-poor environments have been suggested in late-time formation scenarios as we discuss in Section \ref{other_scenarios}.

\subsection{How do mini-Neptunes avoid runaway?}
\label{other_scenarios}

Although this work argues advection from the protoplanetary disks can only play a limited role on envelope formation timescales of mini-Neptunes, the observed population of such $\sim 2-20 M_\oplus$ planets with thin envelopes are still amongst the most common types of exoplanets discovered. While we mention in Section \ref{realistic_alpha_adv} that preventing runaway accretion only with accretion flows may be possible at 0.1 AU, the occurrence rate of mini-Neptunes is constant in log period space beyond $\sim$10 days (i.e., beyond 0.1 AU; see \citealt{Petigura_2018,Wilson_2022}). We thus need to revert back to other scenarios to explain why accretion onto $5-10$ $M_\oplus$ cores does not result in the formation of gas giants. 

Late-stage assembly of the cores remains a promising theory as to how mini-Neptunes can avoid runaway. The delay in the last mass doubling of planetary cores is naturally explained by gas dynamical friction within the initially gas-rich disk preventing the merger of small protocores until the last $\sim$10\% ($\sim 0.1-1$ Myr) of the protoplanetary disk lifetime \citep{Papaloizou_2000,Kominami_2002}. \cite{Lee_2016} showed that the gas-poor conditions within that shorter period of time can be sufficient for core accretion to produce the observed thin envelopes. They also determined that this model is consistent with the degree of disk gas depletion required for mergers of protocores \citep[see also][]{Dawson}. In addition, late-time core assemblies (i.e., mass growth by collisional mergers) are consistent with the observed flat orbital period distribution of these small planets \citep{Lee_2017}, the observed distribution of orbital period ratios in {\it Kepler} multi-planetary systems which feature peaks near (but not in) first-order mean motion resonances \citep[e.g.,][]{Choksi} with the majority of planet pairs being far away from such resonances \citep[e.g.,][]{Izidoro_2017,Izidoro_2021}, as well as the observed intra-system similarity \citep[e.g.,][]{MacDonald_2020,Lammers_2023} whereby planets within a given system have more similar masses and radii compared to system-by-system variations. Therefore, late-time formation of planetary cores remains a likely explanation of the abundance of mini-Neptunes. 

Concurrent pebble accretion may provide additional source of opacity, slowing down the cooling and therefore mass growth of the envelope over the disk lifetime \citep{Ormel_2021}. The ability for heating by additional solid mass accretion to avoid runaway however is limited to short orbital distances ($\lesssim$0.2 AU) and for fine-tuned rate and duration of infall rate (see Figure 12 of \citealt{Ormel_2021}; see also Figure 2 of \citealt{Lee_2015}). Furthermore, sufficiently high atmospheric metallicity can accelerate the runaway process by increasing the gas mean molecular weight (see, e.g., \citealt{Lee_2016}, their Figure 1; see also \citealt{Hori_2011}, \citealt{Venturini_2015} for the specific case of pollution by icy solids, and \citealt{Ormel_2021} for pollution by silicate pebbles).

We note that post formation effects may explain, in part, the thin envelope mass of mini-Neptunes. As shown in Figure \ref{hybrid_vs_radiative_vs_isentropic}, realistically accounting for entropy advection can lead to final GCRs of $\sim$20\% at 0.1 AU. At such separation from the central star ($a \lesssim 0.3$ AU), photoevaporation could play a significant role during the post-formation era in whittling down the final GCR to a value that is consistent with the observed radius and mass \citep[e.g.,][]{Lopez_2012,Owen_Wu_2013,Owen_Wu_2017,Jin_Mordasini_2014}. However, using a hierarchical inference analysis, \cite{Roger_Owen_2021} argue that the typical initial (pre-evaporation) envelope mass fraction of mini-Neptunes must have been $\sim$1--10\% in order to explain the observed radius-period distribution (see their Figure 11) under the theory of photoevaporation \citep[see also][]{Jankovic19} and so we conclude that entropy advection will not be enough even if we consider post-formation photoevaporative mass loss and that further limiting processes such as late-stage assembly is still required.

Alternatively, \cite{Inamdar_2016} demonstrated how giant impacts post-formation (i.e., after the disk gas has completely dissipated away) can result in a loss of approximately half the envelope mass. Furthermore, when accounting for the thermal expansion of the envelope following an impact, \cite{Biersteker_2019} show that small hot planets close to their host stars can even be subjected to total atmospheric loss. Thus, the fully formed dusty envelopes of Figure \ref{hybrid_vs_radiative_vs_isentropic} even at 1 AU could become mini-Neptunes following one or two giant collisions over billion years timescales. It remains unclear however whether the lost material following impact is accreted back to the planet or ejected from the system, unbound from the orbit. Furthermore, N-body calculations report systems that tend to complete the last mergers in gas-{\it free} environments end up with rocky planets ($<$2$R_\oplus$; see Figures 4 and 8 of \cite{MacDonald_2020}) whereas systems with enough material to create more gas-enveloped mini-Neptunes tend to complete the last mergers in gas-poor but not gas-{\it free} environments. We conclude that mass loss by giant impacts may not be so relevant for mini-Neptune populations.

\subsection{Planet population past 1 AU}
\label{predictions}

Currently, little is known about the population of low-mass planets beyond 1 AU as they are mostly detected by transit methods which are only sensitive to orbital periods $\lesssim$300 days \citep[e.g.,][]{Petigura_2018,Hsu_2019,Wilson_2022}. The Nancy Grace Roman Space Telescope (Roman) set to launch in mid-2020 promises to provide some answers with space based microlensing surveys that are uniquely capable of probing masses as low as $\sim$0.01$M_\oplus$ at a few AU \citep{Gaudi_2021,Zhu_Dong_2021}. The amount of gas a planet can accrete depends most sensitively on the mass of the core \citep{Lee_2015,Lee_2019}, a result that is consistent with our calculation (see Figure \ref{GCR_vs_t_all}). Therefore, in this section, we discuss how the current and future measurements of planet masses and mass distribution can be leveraged to distinguish between different formation conditions (e.g., dusty vs.~dust-free accretion, core assembly time, penetration depth of advective flows) at orbital distances from 0.1 to a few AU.

\begin{figure*}
\includegraphics[width=\textwidth]{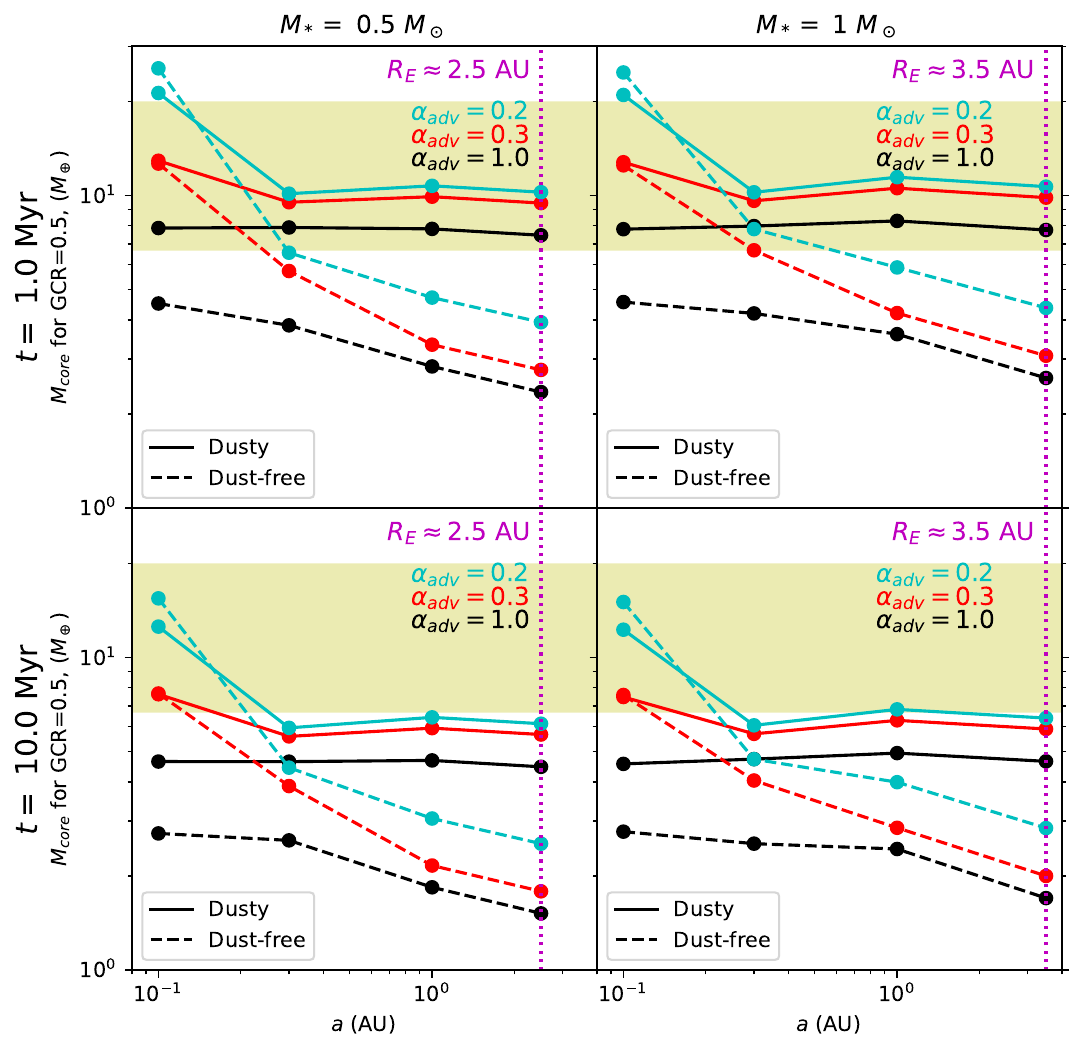}
\caption{Maximal core masses of gas-poor planets (GCR<0.5) expected from the analytic scalings of Section \ref{Scalings} as a function of orbital distance. The top and the bottom rows each correspond to total gas accretion time of 1 Myr, representing late-stage core formation, and to total gas accretion of 10 Myr, representing early core formation. The left and right columns depict M dwarf host stars vs.~solar mass host stars, respectively, with the planetary orbits corresponding to the Einstein ring radius $R_{\rm E} \sim 3.5 {\rm AU} \sqrt{M_*/M_\odot}$ \citep{Suzuki_2018} where planet detection by microlensing is most sensitive indicated in vertical magenta lines. Solid and dashed curves account for dusty and dust-free opacities respectively with the different colors illustrating a range of $\alpha_{\rm adv} =$ 1 (black), 0.3 (red), and 0.2 (cyan). The shaded areas correspond to the estimated core mass of planets that are on the verge of runaway, calculated by taking the measured total mass of planets with radii $\sim$3--8$R_\oplus$ and dividing them by 1.5 accounting for GCR $\sim$0.5; the mass and radii measurements for solar-mass and M dwarf host stars are each taken from \citet[][their Figure 1]{Otegi_2020} and \citet[][their Figure 1]{Luque_2022} respectively. Note that these measurements are only for short orbital periods, specifically for orbital periods less than 35 days for \cite{Luque_2022}.
The maximal core masses are the most distinct between dusty vs.~dust-free opacity at wide orbits.
}
\label{max_core_mass}
\end{figure*}

We focus our attention on sub-Saturns---planets with radii $\sim$3--8$R_\oplus$ which are the boundary population between more gas-poor mini-Neptunes and more gas-rich Jupiters. These planets are unambiguously enveloped by H/He-dominated gas rendering their model-inferred gas-to-core mass ratios more accurate and less subject to the degeneracy between different chemical compositions \citep[e.g.,][]{Petigura_2017}.\footnote{While \cite{Petigura_2017} define the lower limit of sub-Saturn radius as 4$R_\oplus$, we choose slightly smaller 3$R_\oplus$ based on the distinct clusters shown in bulk density-mass space as shown in Figure 1 of \citet{Luque_2022} as well as the clear departure of planets at radii beyond 3$R_\oplus$ from the potential waterworlds and rocky planets in the mass-radius space shown in Figure 1 of \citet{Otegi_2020}.} 

Because the planets formally begin their runaway gas accretion at GCR $\sim$ 0.5 (see Figure \ref{scaling_examples}), we define the mass of the core that reaches this GCR for a given set of formation conditions as the maximal core mass that defines the boundary between those that are gas-poor (and so we can still probe the masses of their cores) vs.~those that are gas-rich (and so their masses are dominated by the gas). We solve for this core mass by inverting GCR scalings of \citet{Lee_2015} modified for the core radius-mass relation of Equation (\ref{R_core}) and the dependence on $\alpha_{\rm adv}$ that we found in Equation (\ref{GCR_prop_to}) which takes the form

\begin{equation}
\label{GCR(alpha_adv,M_core,t)}  
    \mathrm{GCR} = \mathrm{GCR}_0 \Big[\frac{t}{t_0} \Big(\frac{M_{\rm core}}{5M_\oplus} \Big)^{\frac{3}{4}\frac{1+\alpha}{\gamma-1}-1-\frac{\alpha}{4}}\Big]^{\frac{1}{2+\alpha}} \alpha_{\rm adv}^{\theta},
\end{equation}
where we use the values of $\theta$ presented in Figure \ref{GCR_vs_alpha_adv} and $\alpha$ and $\gamma$ are computed at each orbital distance. We fix the initial unrestricted envelope to GCR$_0$ = 0.08 and to the associated computed time $t_0$.

At short orbital distances, our model-predicted maximal core mass can be directly compared with observations using the measured masses and radii of sub-Saturns. From a visual inspection of Figure 1 of \citet{Otegi_2020} and \citet{Luque_2022}, we see that the total masses of such planets range from $\sim$10--30$M_\oplus$ for both solar-type stars and M dwarfs. We divide this total mass by 1.5 to obtain the inferred core mass assuming a typical GCR of 0.5 (shaded region in Figure \ref{max_core_mass}). 

As demonstrated in Figure \ref{max_core_mass}, the maximal core mass under dusty accretion is generally independent of orbital distance (because under dusty accretion, the rcb is set by H$_2$ dissociation and H- opacity which are insensitive to boundary conditions) and larger than that expected from dust-free accretion (because dusty opacity is generally higher than dust-free opacity), in agreement with \citet{Lee_2014}. Under dust-free accretion, the maximal core mass decreases to $\sim$2--4$M_\oplus$ beyond 1 AU, as expected due to diminished opacity farther away from the star where the disk is colder, freezing out the internal modes of gas molecules. We find that at $\lesssim$0.3 AU, both dusty and dust-free accretion provide similar maximal core of $\sim$10$M_\oplus$ in agreement with current observations. At wider orbits, the two opacity models diverge so it is potentially possible to observationally distinguish between the two formation models. Other parameters such as the total time allotted for gas accretion (at least between 1 and 10 Myr) and $\alpha_{\rm adv}$ effect factors of $\lesssim$2 differences and so would be more difficult to disambiguate using planet mass alone.

Current radial velocity measurements are insensitive to these small mass planets beyond $\sim$1 AU. While microlensing studies of \citet{Suzuki_2018} report a potential break in the planet mass distribution at mass ratio $\sim$7$\times$10$^{-5}$ (consistent with $\sim$12$M_\oplus$ assuming 0.5$M_\odot$ host star), the shape of the distribution below this mass ratio is not well constrained and so it remains unclear whether the break is real or not. Theoretically, we would expect such a break to appear at a transition mass between gas-poor and gas-rich planets as the former would more closely track the underlying core mass distribution whereas the latter would track the physics of gas accretion.

In order to provide a more direct prediction of our model that is applicable to microlensing measurements which are more sensitive to M dwarf host stars, we construct the gas accretion evolution of planets around 0.5$M_\odot$ host stars and renormalize our scaling relationship to compute the maximal core mass. We adopt a cooler disk midplane temperature scaling Equation (\ref{temperature_disk}) by $M_\star^{2/7}$ following \cite{Chachan_2023}; in addition, $R_{\rm out}$ would also change accordingly given that $R_H \propto M_\star^{-1/3}$. As shown in the left panels of Figure \ref{max_core_mass}, there is little difference in our maximal core mass between 0.5$M_\odot$ and 1.0$M_\odot$ host stars because the changes in $M_\star$ make minor differences in our boundary conditions. Like the 1.0$M_\odot$ case, we find that the maximal core mass is $\sim$1.5--3.6$M_\oplus$ at the corresponding Einstein ring radius $R_E$ (the planet's orbital distance where microlensing is most sensitive) for dust-free accretion around M dwarfs, significantly smaller than $\sim$4.5--10$M_\oplus$ expected from dusty accretion around the same host stars. 

These two limits can be compared to the measurements of breaks in planet mass distributions from future microlensing surveys such as Roman in order to determine the role of dust opacity (and more generally any physical process that gives rise to orbital-distance-dependent opacities; \citealt{Chachan_Lee_Knutson_2021}) in shaping planet population at orbits beyond $\sim$1 AU. With enough precision, those future observations could also help constrain the typical advection depth as Figure \ref{max_core_mass} shows how different values of $\alpha_{\rm adv}$ give distinct limits on the maximal core mass of gas-poor planets, especially for dust-free opacities.

In this paper, we have focused on the genesis of gas-enveloped planets. Future studies following the post-disk thermal evolution of such planets would quantify the effect of different opacity sources (e.g., dusty vs.~dust-free) on the dependence of final planet sizes on orbital distances. Testing such predictions against future transit missions that can probe small planets ($\lesssim$2$R_\oplus$) out to $\sim$1 AU (e.g., PLAnetary Transits and Oscillations of stars; Plato) {\it and beyond} would offer another avenue of distinguishing between different formation conditions. Making such predictions is a subject of our future work.

\section{Conclusion}
\label{Conclusion}

We have investigated the role of atmospheric recycling in delaying or halting gas accretion by cooling onto planetary cores embedded in gas-rich nebula. Our main findings are summarized as follows:
\begin{enumerate}
    \item When more realistic EOS and opacities are taken into account, along with proper accounting of surface energies in thermal evolution, except for specific scenarios ($\alpha_{\rm adv} < 0.25$, 0.1 AU), advection alone is insufficient to produce mini-Neptunes because even if runaway may be avoided, the final envelope mass fraction of mini-Neptune mass planets would be $\gtrsim$0.2, too large to explain their measured masses and radii. Further limiting process such as late-time core assembly is required.

    \item At close-in distances (0.1 AU), the outer advective layer is expected to be isentropic with the disk while at large distances (1 AU), this outer layer is found to be more radiative for dust-free opacity so the effect of entropy advection is most apparent at 0.1 AU and less so at larger orbital separations.

    \item The dependence of the final gas-to-core mass ratio (GCR) on the penetration depth of advective flow generally strengthens for deeper flows (lower $\alpha_{\rm adv}$) and at close-in orbital distances. The critical penetration depth $\alpha_{\rm adv}R_{\rm out}$ where the behavior of the GCR-$\alpha_{\rm adv}$ scaling relationship changes is at the locations of initial radiative-convective boundaries in corresponding unrestricted ($\alpha_{\rm adv} = 1$) envelopes with all else equal.

    \item The critical core mass that separates gas-poor vs.~gas-rich planets is most sensitively determined by the nature of opacity (dusty vs.~dust-free) more so than the core assembly times or $\alpha_{\rm adv}$ and this difference in the critical core mass is more pronounced at larger orbital distances beyond $\sim$1 AU. Such a deviation presents a timely and unique opportunity to leverage any observable break in the planet mass distribution from e.g., microlensing survey with Roman to distinguish between dusty vs.~dust-free gas accretion.

\end{enumerate}

Our findings highlight the importance of EOS and opacities in regulating the thermal evolution of planetary envelopes which are the key difference between our calculations than that of e.g., \citet{Ali_Dib_2020} that lead to different conclusions. Future three-dimensional hydrodynamic simulations that employ realistic EOS and opacities over a wide range of orbital distances would be welcome to verify the 1-D calculations showcased in this work.

\section*{}

We thank the referee for a helpful report. We are indebted to Jason Ferguson for extending and sharing his opacity tables, and we thank James Owen and Zhaohuan Zhu for their insight and helpful discussions. We also thank Eugene Chiang and Andrew Cumming for their feedback on the earlier versions of this paper. V.S. acknowledges the support of the Natural Sciences and Engineering Research Council of Canada (NSERC), of le Fonds de recherche du Québec – Nature et technologies (FRQNT) and of the Trottier Institute for Research on Exoplanets (iREx) under the Trottier Excellence Grant for Summer Interns. E.J.L. gratefully acknowledges support by NSERC, by FRQNT, by the Trottier Space Institute, and by the William Dawson Scholarship from McGill University.

\appendix

\counterwithin{figure}{section}

\section{Parameter study}

\begin{figure*}[h!]
\subfigure[Dusty]{\includegraphics[width=8.45cm]{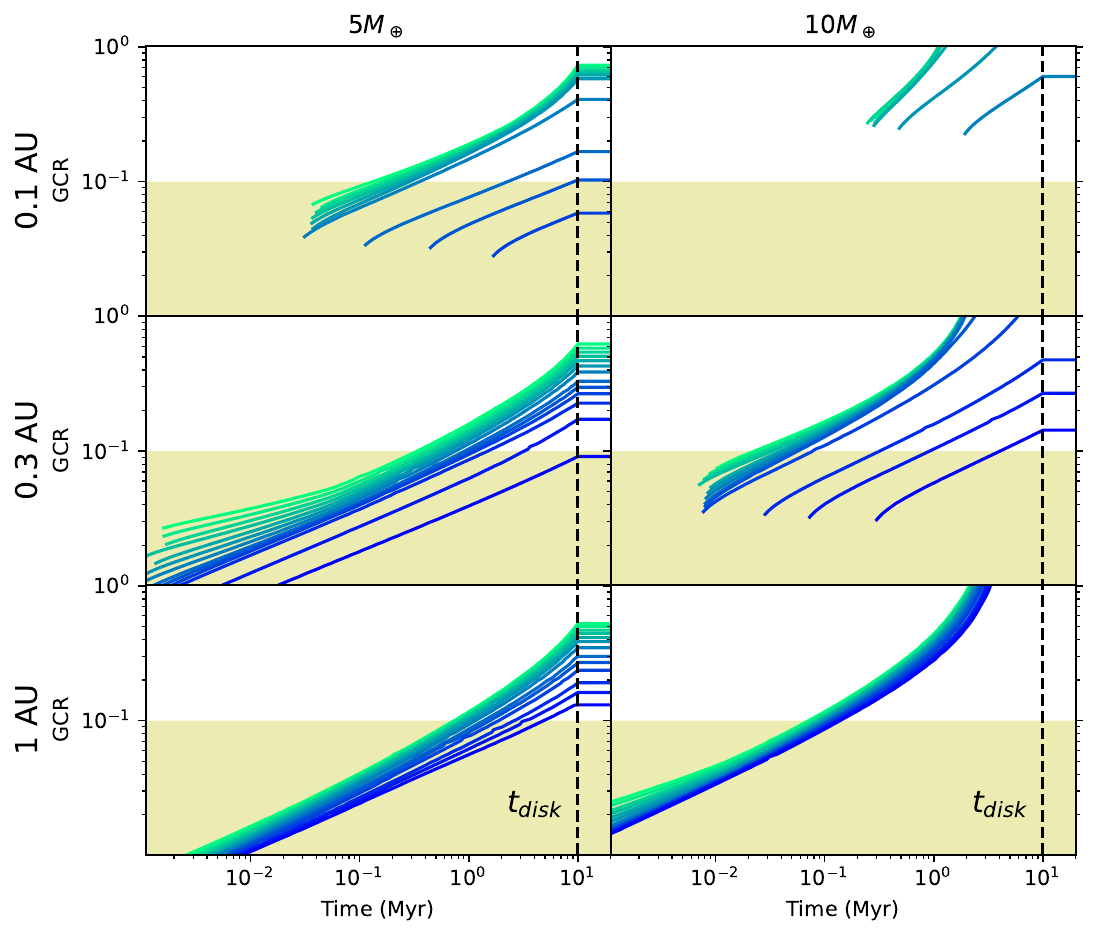}}
\subfigure[Dust-free]{\includegraphics[width=9.55cm]{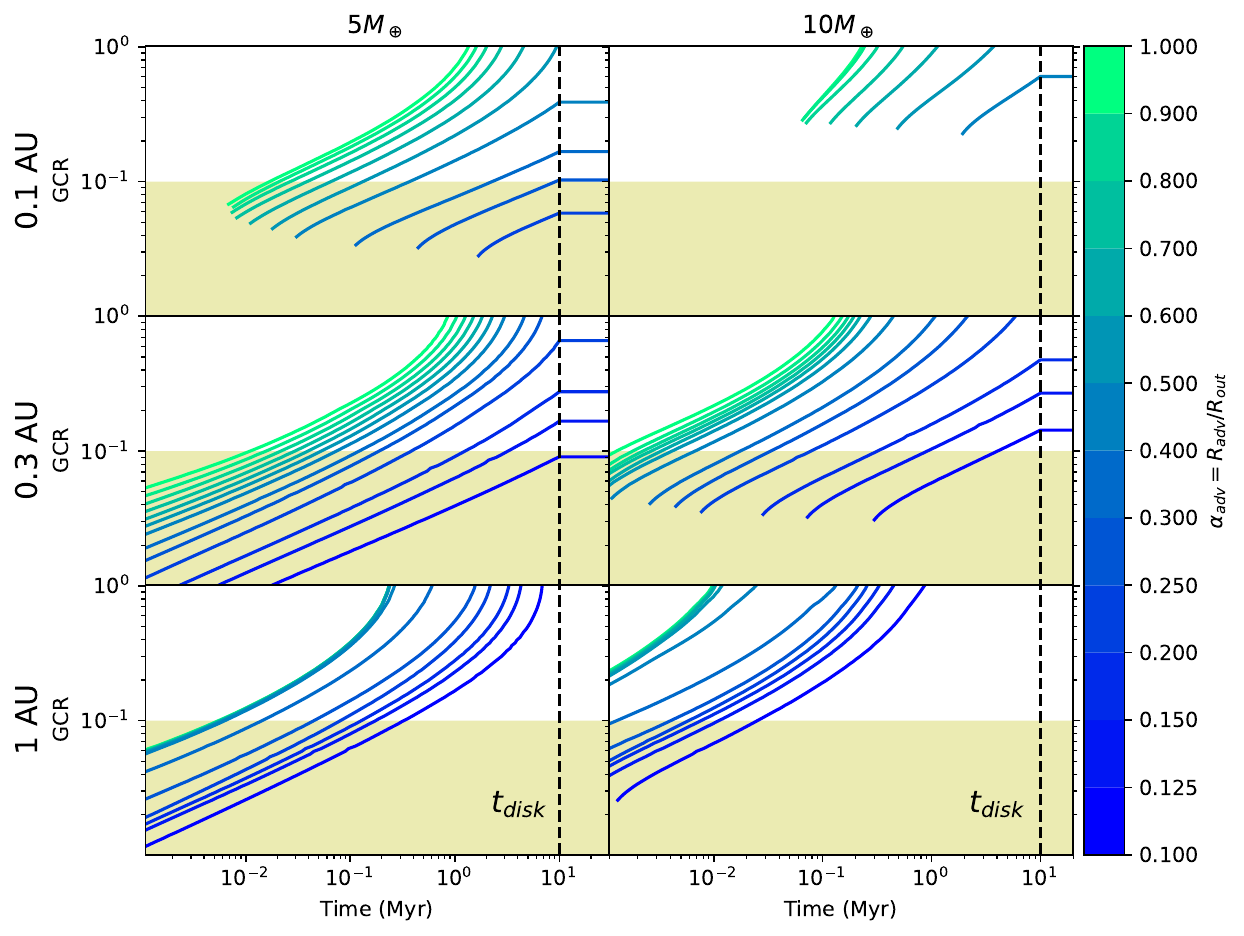}}
\caption{Parameter study of the effect of advective flows on the time evolution of the GCR of a core embedded in a gas-rich nebula under different boundary conditions. The left and right figures account for dusty and dust-free opacities respectively. From top to bottom, the orbital distance is varied from 0.1 AU to 0.3 AU to 1 AU. The first column of each figure considers a $5M_\oplus$ core while the second one a $10M_\oplus$ core. We show a range of 13 values of entropy advection depth ($\alpha_{\rm adv} = R_{\rm adv}/R_{\rm out}$) decreasing from 1 to 0.1 indicated by the color bar accompanying the figure. The target range for mini-Neptunes envelope mass fraction (shaded region) and the protoplanetary disk lifetime (dashed vertical line) are indicated as references.}
\label{GCR_vs_t_all}
\end{figure*}

A gallery of envelope mass growth profiles across our entire parameter space is presented in Figure \ref{GCR_vs_t_all}. We recover the strong dependence of GCR on the core mass finding a more massive 10$M_\oplus$ core to undergo runaway accretion in unrestricted envelopes \cite[e.g.,][]{Mizuno,Stevenson,Pollack,Ikoma}. While our model in those cases is limited to high GCR at 0.1 AU as envelopes are fully convective at early times, runaway is already unavoidable at 1 AU for formation in gas-rich environments, showcasing the need for further limiting process (see Section \ref{other_scenarios}) is particularly dire for massive cores.

\bibliography{References}
\bibliographystyle{aasjournal}

\end{document}